\documentclass[prd,twocolumn,superscriptaddress,showpacs,nofootinbib,
preprintnumbers]{revtex4}

\usepackage{amsmath}
\usepackage{amsfonts}
\usepackage{graphicx}
\usepackage{dcolumn}

\def\be{\begin{equation}}
\def\ee{\end{equation}}
\def\ba{\begin{eqnarray}}
\def\ea{\end{eqnarray}}
\def\bs{\begin{subequations}}
\def\es{\end{subequations}}
\def\R{{\cal R}}
\def\Sc{{\cal S}}

\newcommand{\rd}{{\rm d}}

\usepackage{color}

\begin{document}

\title{Prospects of inflation in delicate D-brane cosmology}

\author{Sudhakar Panda}
\affiliation{Harish-Chandra Research Institute, Chhatnag Road,
Jhusi, Allahabad-211019, India}
\email{panda@mri.ernet.in}

\author{M.~Sami}
\affiliation{Centre of Theoretical Physics, Jamia Millia Islamia,
New Delhi-110025, India} \email{sami@jamia-physics.net}

\author{Shinji Tsujikawa}
\affiliation{Department of Physics, Gunma National College of
Technology, Gunma 371-8530, Japan}
\email{shinji@nat.gunma-ct.ac.jp}

\date{\today}

\begin{abstract}

We study D-brane inflation in a warped conifold background
that includes brane-position dependent corrections for the nonperturbative 
superpotential. Instead of stabilizing the volume modulus $\chi$ at 
instantaneous minima of the potential 
and studying the inflation dynamics with an effective 
single field (radial distance between a brane and an anti-brane) $\phi$,
we investigate the multi-field inflation scenario involving these two 
fields. The two-field dynamics with the potential
$V(\phi, \chi)$ in this model is significantly different from
the effective single-field description in terms of the field $\phi$ when
the field $\chi$ is integrated out.
The latter picture underestimates the total number of
e-foldings even by one order of magnitude.
We show that a correct single-field description is provided
by a field $\psi$ obtained from a rotation in the two-field space along 
the background trajectory. This model can give a large number of
e-foldings required to solve flatness and horizon problems
at the expense of fine-tunings of model parameters.
We also estimate the spectra of density perturbations and
show that the slow-roll parameter
$\eta_{\psi \psi}=M_{\rm pl}^2 V_{,\psi \psi}/V$
in terms of the rotated field $\psi$ determines the
spectral index of scalar metric perturbations.
We find that it is generally difficult to satisfy, simultaneously, both 
constraints
of the spectral index and the COBE normalization, while the
tensor to scalar ratio is sufficiently small to match
with observations.

\end{abstract}

\maketitle

\section{Introduction}

Modern Cosmology witnessed a revolution in 1980 with the advent of
cosmological inflation \cite{inflation}.
The paradigm has stood the test of theoretical and
observational challenges in the past two decades \cite{Spergel1,Spergel2}.
In spite of its cosmological successes to solve horizon and flatness problems,
it still remains a paradigm in search of a viable theoretical model.
It is, therefore, not surprising that efforts are underway to derive 
inflationary models from string theory, a consistent quantum
field theory around the Planck's scale. The discovery of nonperturbative
solutions, in string theory, called D-branes \cite{Pol} has given rise to new
hopes in this direction.

The D-brane cosmology is a subject of intense study at present. It
ranges from inflation on a non-BPS brane because of tachyon
condensation to inflation due to the motion of a D3-brane towards an
anti-D3-brane \cite{sen,linde,kallosh,lindeD}. Efforts also have
been made to study inflation due to geometric tachyon arising from
the motion of a probe brane in the background of a stack of either
NS5-branes or the dual D5-branes \cite{GTach} . These models are
constructed in the framework of effective field theory and assume an
underlying mechanism for the stabilization of various moduli fields.

An important step towards a realistic inflationary model in string
theory emerged from the realization that background fluxes can
stabilize most of the modulus fields.
As shown in Ref.~\cite{GKP}, the fluxes in a warped compactification, 
using a
Klebanov-Strassler (KS) throat \cite{KS}, can stabilize the dilaton
and complex structure moduli of type IIB string theory compactified
on an orientifold of a Calabi-Yau threefold. In fact it was
demonstrated in Ref.~\cite{KKLT} that all the closed string moduli
can be stabilized by a combination of fluxes and nonperturbative
effects. The nonperturbative effects are mainly responsible for
stabilizing the K\"ahler moduli; they arise, via gauge dynamics of either 
an Euclidean D3-brane or from a stack of
$n$  D7-branes wrapping super-symmetrically a four cycle in
the warped throat. The warped volume of the four cycle controls the
magnitude of the nonperturbative effect since it affects the gauge
coupling on the D7-branes wrapping this four cycle.

The above formalism could lead to the construction of a realistic
inflationary model \cite{KKLMMT}, in string theory, which is built
upon the compactification data (see also Refs.~\cite{dbpapers}).
The inflaton potential can be
obtained by performing string theoretic computations involving the
details of the compactification scheme. In this setup inflation
is realized by the motion of a D3-brane towards a distant static
anti-D3-brane, placed at the tip of the throat. The position of the
moving brane in the compactification manifold is identified with the
inflaton field. To be more precise, the location of the mobile brane
can be labeled with five angular coordinates and one radial
coordinate, $r$. The canonical inflaton field $\phi$ is expressed by
a constant re-scaling of this radial coordinate.

In Ref.~\cite{Bau1}, the embedding of the D7-branes as given in
\cite{Kup} was considered. It was assumed that at least one of the
four-cycles carrying the nonperturbative effects descend down a
finite distance into the warped throat. It was then shown that the
presence of a D3-brane gives rise to a perturbation to the warp
factor affecting a correction to the warped four cycle volume.
Moreover, this correction depends on the position of the D3-brane
and thus the superpotential for the nonperturbative effect gets
corrected by an overall position-dependent factor. The total
potential, that the mobile D3-brane experiences, is the sum of the
potential (F-term) coming from the superpotential  and the usual
D-term potential contributed by the interaction between the D3-brane
and the anti-D3-brane. When the corrections to the
nonperturbative superpotential is taken into account, the issue of volume 
modulus
stabilization needs to be re-analyzed. This has been carried out in
Refs.~\cite{Bau2,Bau3,KP} and  the viability of inflation was
investigated in this modified scenario.

The model in Ref.~\cite{Bau2,Bau3} is described by a two-field
potential $V(\phi,\sigma)$ in terms of the inflaton $\phi$ and
the volume modulus $\sigma$. If the mass of the modulus
is much larger than the Hubble rate, the field $\sigma$
approximately evolves along the instantaneous minima determined
by the condition $V_{,\sigma}=0$. One can obtain an effective
single-field potential with respect to $\phi$ by integrating out
$\sigma$ in this way. However, we need to be careful for the fact that the 
actual trajectory
is determined by the direction along the velocity of the
fields. We shall introduce a new rotated field $\psi$ along the
background trajectory and show that
the usual single-field description holds for $\psi$ and not for $\phi$.
As a result the single-field description in terms of $\phi$ underestimates
the total amount of e-foldings even by one order of magnitude.
This reflects that fact that the amount of inflation is sensitive to
the slight change of the potential and also of model parameters.

We shall also calculate the spectra of density perturbations generated
in this model. Again the spectral index $n_{\cal R}$ is determined by
the slow-roll parameter $\eta_{\psi \psi}=M_{\rm pl}^2 V_{,\psi \psi}/V$
instead of $\eta_{\phi \phi}=M_{\rm pl}^2 V_{,\phi \phi}/V$.
Thus we show that the correct single-field description in terms of
$\psi$ is crucially important to study the perturbation spectra
as well as the background dynamics.

\section{The D3-brane potential}

In this section we review the derivation of the scalar potential
\cite{Bau2, Bau3} on a mobile D3-brane. The fluxes for the
compactification of type IIB string theory on an orientifold of
Calabi-Yau theory are chosen such that the internal space has a
warped throat region. For example, the local geometry is taken to be
the warped deformed conifold which is a subspace of complex
dimension three in a four dimensional complex space defined by the
constraint $\sum_{i=1}^4 z_i^2 = \epsilon^2$ where $z_i$ are
coordinates in the four dimensional complex space and $\epsilon$ is
real and corresponds to the deformation parameter. When the region
of relevance for the D-brane inflation is chosen to lie far from the
tip of the throat, the deformed parameter can be neglected. We can
then choose $z_\alpha = (z_1, z_2, z_3)$ as the three independent
complex coordinates for the position of the D3-brane (open string
moduli) and use the conifold constraint to express $z_4$ in terms of
these three coordinates. The throat can be glued into a compact
space, which is assumed to have a single K\"ahler modulus $\rho$.

One can take the K\"ahler potential to be
\be
\label{kapot}
k=\frac{3}{2} \left(\sum_{i=1}^4 |z_i|^2\right)^{2/3}~=\frac{3}{2}
r^2~={\hat r}^2\,,
\ee
so that the K\"ahler metric on the conifold, $k_{\alpha, {\bar \beta}}
\equiv \partial_\alpha \partial_{\bar\beta} k$, is Ricci-flat. Thus the
metric of the deformed conifold can be written as
$\rd s_6^2 = \rd{\hat r}^2 +
{\hat r}^2 \rd s^2_{T^{1,1}}$, where $\rd s^2_{T^{1,1}}$ is the metric on the
Einstein space $T^{1,1}$, which can be expressed in terms of five angular
coordinates. These angular coordinates and the real radial coordinate $r$
are the basis for the complex coordinates $z_i$,
see Ref.~\cite{Bau3} for details.

The $N = 1$ supergravity F-term scalar potential involving the
DeWolfe-Giddings K\"ahler potential ${\cal K}$
and the super potential $W$ is given by
\be \label{knew}
V_F=e^{\kappa^2 {\cal K}} \left[ D_\Sigma W {\cal
K}^{\Sigma{\bar \Gamma}} {\bar D_\Gamma W}
- 3 \kappa^2 W {\bar W} \right]\,,
\ee
where $D_\Sigma W = \partial_\Sigma W + \kappa^2 (\partial_\Sigma
{\cal K}) W$, $\left\{Z^\Sigma\right\} \equiv \left\{\rho, z_\alpha;
\alpha = 1, 2, 3\right\}$ and $\kappa^2 = M_{\rm pl}^{-2} \equiv 8\pi G$.
The total K\"ahler potential ${\cal K}$ depends upon $\rho$ and the
open string moduli $z_\alpha$, and is given by \cite{DG}
\ba
\label{kahler}
{\cal K} (\rho,{\bar \rho}, z_\alpha, {\bar z_\alpha})
&=& -3M_{\rm pl}^2\,\,{\rm ln} [\rho + {\bar \rho} - \gamma k (z_\alpha, {\bar
z_\alpha})]\nonumber \\
&\equiv& - 3M_{\rm pl}^2\,\,{\rm ln} U\,,
\ea
where $k (z_\alpha, {\bar z_\alpha})$ is the little K\"ahler
potential, as defined in Eq.~(\ref{kapot}), for the metric on the
conifold. The parameter $\gamma$
is proportional to the ratio of warped volumes of 
four cycle (before the D3-brane enters the throat) and the three-fold.
In the expression (\ref{knew}) for the scalar potential, ${\cal
K}^{\Sigma {\bar \Gamma}}$ is the inverse K\"ahler metric which can be
derived from the K\"ahler potential ${\cal K}$. This leads to the
following result
\begin{widetext}
\ba
\label{ftpot1}
V_F (\rho, z_\alpha) &=& \frac{\kappa^2}{3 U^2}
\Biggl[ \left\{ \rho + {\bar \rho}
+ \gamma (k,_\alpha k^{\alpha{\bar \beta}}
k,_{{\bar \beta}} - k )\right\}
|W,_\rho|^2 - 3 ({\bar W} W,_{\rho} + c.c) \nonumber \\
& &~~~~~~~+(k^{\alpha{\bar \beta}} k,_{{\bar \beta}}
{\bar W,_\rho} W,_\alpha +
c.c) + \frac{1}{\gamma} k^{\alpha{\bar \beta}} W,_\alpha
{\bar W,_\beta} \Biggr].
\ea
\end{widetext}
The first line in Eq.~(\ref{ftpot1}) is the standard
Kachru-Kallosh-Linde-Trivedi (KKLT) F-term
potential \cite{KKLT}, while the rest of terms owe their existence
entirely to the dependence of the nonperturbative superpotential on
the open string moduli \cite{Bau1} (see also Refs.~\cite{Berg}).

This expression simplifies further, when we make use of the conifold
metric (and its inverse) computed from the K\"ahler potential
given in Eq.~(\ref{kapot}), to
\begin{widetext}
\ba
\label{ftpot2}
V_F (\rho, z_\alpha)= \frac{\kappa^2}{3 U^2} \left[( \rho + {\bar \rho})
|W,_\rho|^2 - 3 ({\bar W} W,_{\rho} + c.c. )
+\frac {3}{2}( z_\alpha {\bar W,_\rho} W,_\alpha + c.c. ) +
\frac{1}{\gamma} r \left( \delta^{\alpha \beta} + \frac {1}{2} \frac
{z_\alpha{\bar z_\beta}}{r^3} - \frac {z_\beta {\bar z_\alpha}}{r^3}
\right) W,_\alpha {\bar W,_\beta}\right].
\ea
\end{widetext}
Note that the expression for
$U$, now, is simply $U (\rho, r)=\rho + {\bar \rho} - (3/2)\gamma
r^2$. To obtain the full potential, we add to (\ref{ftpot2}), the
contribution of an anti-D3-brane at the tip of the conifold,
including its Coulomb interaction with
the moving D3-brane \cite{KKLT}:
\be
\label{dpot} V_D (\rho, r)=
\frac {D_0}{U^2 (\rho, r)} \left[ 1
- \frac {3 D_0}{16 \pi^2 ( T_3 r^2 )^2}\right]\,,
\ee
where $D_0 \equiv 2 h_0^{-1} T_3$ is twice the warped D3-brane
tension at the tip of the throat.
The total potential on the mobile brane is thus given by
$V=V_F (\rho, z_\alpha)+V_D (\rho, r)$.

The superpotential $W = W_0 + W_{\rm np}$ where $W_0$ is the
Gukov-Vafa-Witten flux super potential \cite{GVE} and
$W_{\rm np} = A (z_\alpha) e^{- b \rho}$ is the contribution from the
nonperturbative effects. Here the factor, $b \equiv 2\pi/n$,
arises from gauge dynamics on a stack of $n$ number of D7-branes.
These D7-branes wrap a four-cycle in the warped throat preserving
supersymmetry which is specified by the embedding equation
$f(z_\alpha) =0$. The presence of the D3-brane also leads to a
perturbation to the warp factor which results in a correction to the
warped four-cycle volume. This correction has been found to be
D3-brane position dependent and, in fact, is responsible for the
pre-factor $A (z_\alpha)$ in $W_{\rm np}$ \cite{GM}.
In Ref.~\cite{Bau1} this position dependence is found to be
\be
\label{Az}
A (z_\alpha)=A_0 \left (\frac
{f(z_\alpha)}{f(0)}\right )^{1/n}\,,
\ee
where $A_0$ is related to the
threshold corrections which depend on complex structure moduli and
for us it is just a constant parameter as these are already
stabilized by the flux background. 
Note that the relation (\ref{Az}) can be derived generically 
without specifying the embeddings in the warped 
deformed conifold \cite{Koe}. The potential,
$V (\rho, r, z_i)$, thus is a complicated function of the
K\"ahler modulus, the radial
and the five angular coordinates of the mobile D3-brane. However,
the angular coordinates can be integrated out as mentioned below.

Choosing the holomorphic embedding of Ref.~\cite{Kup}, given by the equation
$f(z_1)=\mu - z_1=0$, we have $A(z_1)=A_0 (1 -z_1/\mu)^{1/n}$.
Using this and also setting $\rho = \sigma + i \tau$ in Eq.~(\ref{ftpot2})
we find
\begin{widetext}
\ba
\label{ftpot3}
V_F = \frac{\kappa^2 b |A|^2 e^{-2 b \sigma}}{3 U^2}
\left[ 2 b \sigma + 6
+ 6 W_0 e^{b \sigma} {\rm Re} \left( \frac {e^{i b \tau}}{A}\right) +
\frac {3}{2n} \frac {\mu (z_1 + {\bar z_1}) - 2 |z_1|^2}{|\mu -
z_1|^2} +  \frac{r}{b \gamma} \left ( 1 - \frac {|z_1|^2}{2
r^3}\right) \frac {1}{n^2 |\mu - z_1|^2}\right]\,.
\ea
\end{widetext}
Note that this potential has only one term that depends on $\tau$.
The potential for $\tau$ is minimized when this term takes the
smallest possible value. Since the coefficient of this term contains
$W_0$ which is negative, integrating out this field amounts to
replacing $e^{i b \tau}/A$ by $|A|^{-1}$. Similarly, the five
angular coordinates, which describe the position of the D3-brane on
the base of the throat, are periodic coordinates on a compact space.
Thus, the potential in these fields is either constant or has
discrete minima for some fixed values of these five fields. Since
the radial motion of the brane is of special interest, we can focus
on the stable trajectories in the angular directions which minimize
the potential.

In the case of the embedding $f (z_1) = 0$, the stable
minima in angular directions occur only for trajectories that obey
$z_1 = - r^{3/2}/\sqrt {2}$ \cite{Bau3}.
From the DBI action of D3-brane, one can read out that the canonical field 
is $\phi =\sqrt {3T_3/2}r$, which corresponds to the 
approximation (when compared to the result obtained from the K\"ahler 
potential) that 
$\rho+\bar{\rho} \gg (\gamma/T_3)\phi^2$ and
that the field $\sigma$ does not change much.
Using this expression for $z_1$ in Eq.~(\ref{ftpot3})
and defining the minimal radial
coordinate (position) of the D7-brane embedding
to be $r_\mu^3 \equiv 2 \mu^2$ i.e.,
$\phi_\mu^2 = (3/2) T_3 (2\mu^2)^{2/3}$,
we can write the full potential, involving only two
fields, the field $\phi$ and the volume modulus $\sigma$,
in the following form
\begin{widetext}
\ba \label{potential}
V(\phi, \sigma) &=&
\frac{b|A_0|^2}{3M_{\rm
pl}^2} \frac{e^{-2b \sigma}}{U^2(\phi,\sigma)} g^{2/n}(\phi) \left[
2b\sigma  +6-6e^{b \sigma} \frac{|W_0|}{|A_0|}\frac{1}{g^{1/n}(\phi)}
+\frac{3}{n} \left\{c \frac{\phi}{\phi_\mu} -\left(
\frac{\phi}{\phi_\mu} \right)^{3/2} -\left( \frac{\phi}{\phi_\mu}
\right)^{3}
\right\} \frac{1}{g^2(\phi)} \right] \nonumber \\
& &+\frac{D(\phi)}{U^2(\phi,\sigma)}\,, \ea
\end{widetext}
where
\ba U(\phi, \sigma) &=& 2\sigma-
\frac{\gamma}{T_3} \phi^2 \,,\\
g(\phi) &=& 1+\left(
\frac{\phi}{\phi_\mu} \right)^{3/2}\,,\\
D(\phi) &=& D_0 \left( 1-\frac{27D_0}
{64\pi^2 \phi^4} \right)\,,
\ea
and $c=1/(6\pi \gamma T_3 \phi_\mu^2)$.

We will use this form of the two-field potential in our analysis in
the next section keeping in mind that $A_0$ and $W_0$ have the
dimension $[{\rm mass}]^3$ while for $D_0$ it is $[{\rm mass}]^4$.
It should be noted that the field $\sigma$ is not yet in the
canonical form. {}From the K\"ahler potential (\ref{kahler}), we find
that the kinetic term, $-K_{\rho \bar{\rho}} \partial_{\mu}\rho
\partial^{\mu} \bar{\rho}$, becomes canonical (at the tip of the 
throat, which coincides with the same approximation made for the 
canonical field $\phi$) if we consider the field $\chi$
defined by 
\be
\frac{\chi}{M_{\rm pl}}=\sqrt{\frac{3}{2}}\,{\rm ln}\,\sigma\,.
\ee
In what follows we shall examine the viability of inflation in the
frame work of two-field dynamics described by the potential $V(\phi,
\chi)$. This potential amounts to replacing $\sigma$ in
Eq.~(\ref{potential}) by $\exp(\sqrt{2/3}\,\chi/M_{\rm pl})$.

\section{Background evolution}

In this section we discuss the dynamics of inflation induced
by the D3-brane potential (\ref{potential}) but with the canonical field 
$\chi$ replacing the field $\sigma$.
In the flat Friedmann-Robertson-Walker (FRW) metric with
a scale factor $a$ the equations of motion are given by
\ba
& & \dot{H}=-\frac{1}{2M_{\rm pl}^2}
( \dot{\phi}^2+ \dot{\chi}^2 )\,, \\
\label{phiequ}
& & \ddot{\phi}+3H\dot{\phi}+V_{,\phi}=0\,, \\
\label{chiequ}
& & \ddot{\chi}+3H\dot{\chi}+V_{,\chi}=0\,,
\ea
together with the constraint equation
\ba
3H^2=\frac{1}{M_{\rm pl}^2}
\left[ \frac12 \dot{\phi}^2+ \frac12
\dot{\chi}^2+V(\phi, \chi) \right]\,,
\ea
where a dot represents a derivative with respect to
cosmic time $t$ and $H \equiv \dot{a}/a$ is
the Hubble parameter.
For later convenience
we introduce the following slow-roll parameters
\ba
& &\epsilon_\phi=\frac{M_{\rm pl}^2}{2}
\left( \frac{V_{,\phi}}{V} \right)^2\,,\quad
\epsilon_\chi=\frac{M_{\rm pl}^2}{2}
\left( \frac{V_{,\chi}}{V} \right)^2\,,
\nonumber \\
& & \eta_{\phi \phi}=M_{\rm pl}^2
\frac{V_{,\phi \phi}}{V}\,, \quad
\eta_{\chi \chi}=M_{\rm pl}^2
\frac{V_{,\chi \chi}}{V}\,, \quad
\eta_{\phi \chi}=M_{\rm pl}^2
\frac{V_{,\phi \chi}}{V}. \nonumber \\
\ea

The squared masses of the fields $\phi$ and $\chi$ are defined by
$m_{\phi \phi}^2 \equiv V_{,\phi \phi}$ and
$m_{\chi \chi}^2 \equiv V_{,\chi \chi}$, respectively.
Since the Hubble parameter approximately satisfies
the relation $3H^2 \simeq V/M_{\rm pl}^2$ during inflation,
the slow-roll parameters $\eta_{\phi \phi}$
and $\eta_{\chi \chi}$ are given by
\ba
\eta_{\phi \phi} \simeq \frac{m_\phi^2}{3H^2}\,,\quad
\eta_{\chi \chi} \simeq \frac{m_\chi^2}{3H^2}\,.
\ea
Since we are considering the case in which the field $\phi$
plays the role of the inflaton, we require that
$\eta_{\phi \phi}$ is not much larger than unity.
Meanwhile the modulus field $\chi$ can be heavy
($\eta_{\chi\chi} \gtrsim 1$) or light ($\eta_{\chi\chi} \lesssim 1$)
depending on the model parameters.
The latter corresponds to the situation in which two stages of
inflation can be realized.

When the $\chi$ mass is much larger than the
Hubble rate ($\eta_{\chi \chi} \gg 1$),
the field $\chi$ rapidly rolls down toward instantaneous
minima of the potential given by $V_{,\chi}=0$.
In this case it was shown in Ref.~\cite{Bau2,Bau3} that the evolution of
the non-canonical field $\sigma=\exp(\sqrt{2/3}\,\chi/M_{\rm pl})$
is approximately described by
\ba
\label{chire}
\sigma_* \approx \sigma_0 \left[ 1+c_{3/2}
\left( \frac{\phi}{\phi_{\mu}} \right)^{3/2} \right]\,.
\ea
We note that this relation was also derived in Ref.~\cite{KP}.
Here $\sigma_0=3\gamma M_{\rm pl}^2/T_3$ and
$c_{3/2}=\left(1-1/2\omega_F\right)/n\omega_F$, where
$\omega_F$ satisfies the relation
$3e^{\omega_F}|W_0|/|A_0|=2\omega_F+3$.
The condition, $\omega_F \gg 1$,  is assumed
to reach the expression (\ref{chire}).
We also note that the approximation (\ref{chire})
is not accurate in the large $\phi/\phi_{\mu}$
region close to 1.

\begin{figure}
\includegraphics[height=3.1in,width=3.1in]{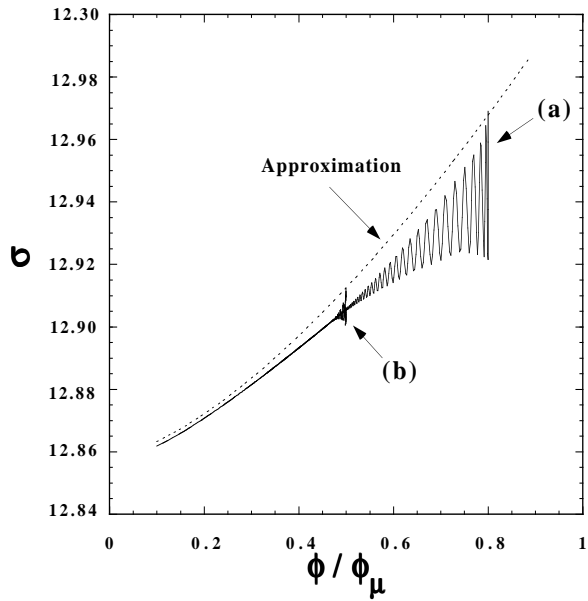}
\caption{\label{fig1} The trajectory of the scalar fields in the
$(\phi,\sigma)$ plane for the model
parameters $n=8$, $A_0=1$, $\omega_0\equiv b\sigma_0=10.1$,
$\omega_F=9.9951$, $W_0=3.496 \times 10^{-4}$,
$D_0=1.215 \times 10^{-8}$ and $\phi_{\mu}=0.25$.
The curves (a) and (b) correspond to the trajectories derived
by solving the background equations numerically for
the initial conditions $\phi_i/\phi_\mu=0.8$ and
$\phi_i/\phi_\mu=0.5$, respectively. The initial values of
the field $\chi$ are chosen to satisfy the relation
(\ref{chire}) with $\sigma_0=\omega_0/b=12.8597$.  
We also plot the approximate trajectory
(\ref{chire}).}
\end{figure}

In Fig.~\ref{fig1} we illustrate a typical example for the trajectory
of the two fields in the $(\phi, \sigma)$ plane.
The case (a) corresponds to the initial conditions $\phi_i/\phi_\mu=0.8$
and $\sigma_i$ satisfying Eq.~(\ref{chire}).
Since the approximation (\ref{chire}) is not accurate in the large
$\phi/\phi_\mu$ region, the initial position of the field $\sigma$
does not exactly match with the local minimum of
the potential ($V_{,\chi}=0$).
The field $\sigma$ quickly oscillates around instantaneous
minima of the potential at the initial stage.
In this case the system enters the slow-roll inflation stage
around $\phi/\phi_\mu \lesssim 0.5$ after the field $\chi$
is almost stabilized at the instantaneous minima.
If we choose smaller initial $\phi/\phi_\mu$ (such as
the case (b) in Fig.~\ref{fig1}), the period of the oscillation of $\chi$
becomes very short.
Figure \ref{fig1} shows that the accuracy of the approximation
(\ref{chire}) becomes better for smaller values of  $\phi/\phi_\mu$.

\begin{figure}
\begin{centering}
\includegraphics[height=3.1in,width=3.1in]{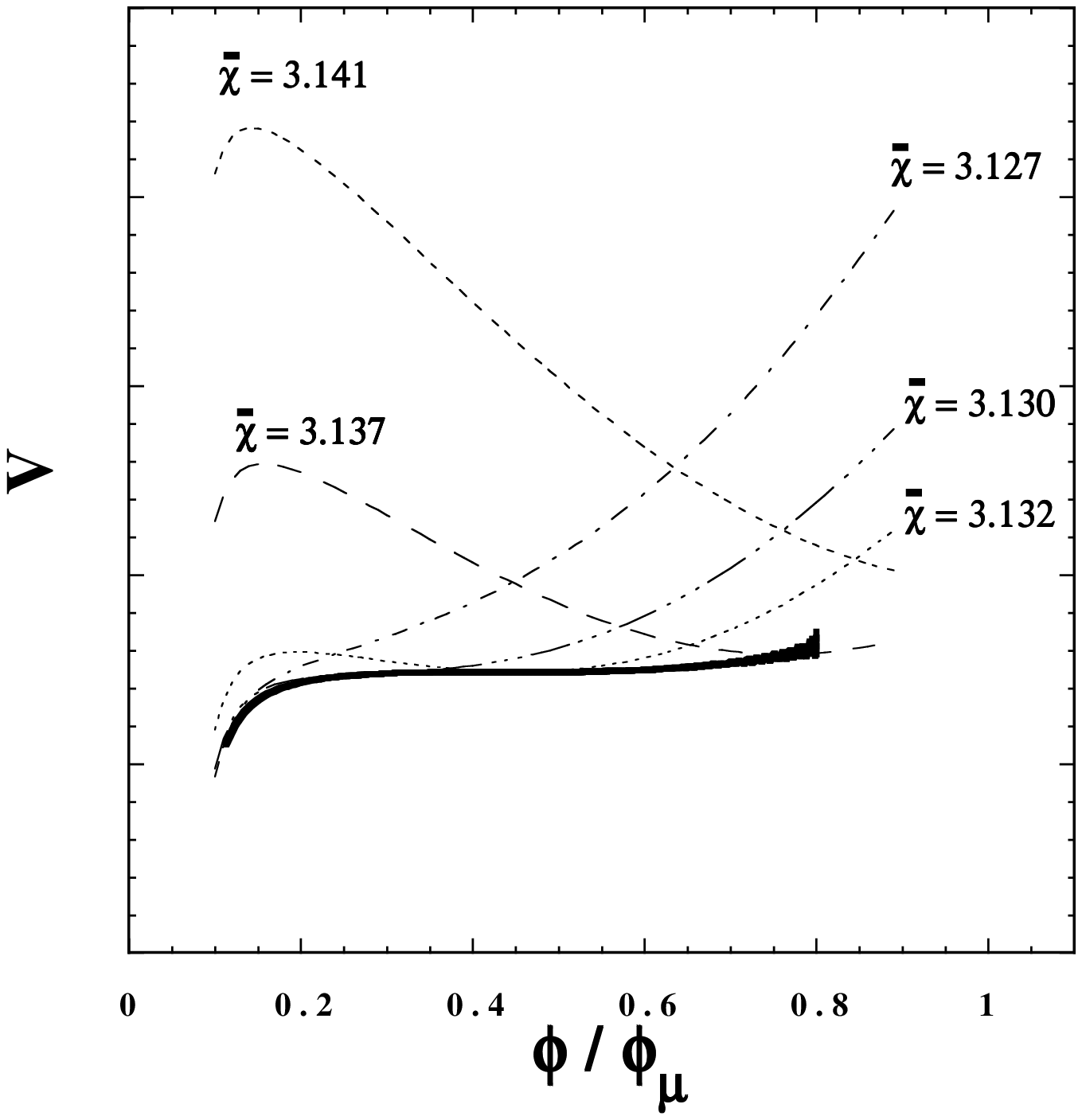}
\includegraphics[height=3.3in,width=3.3in]{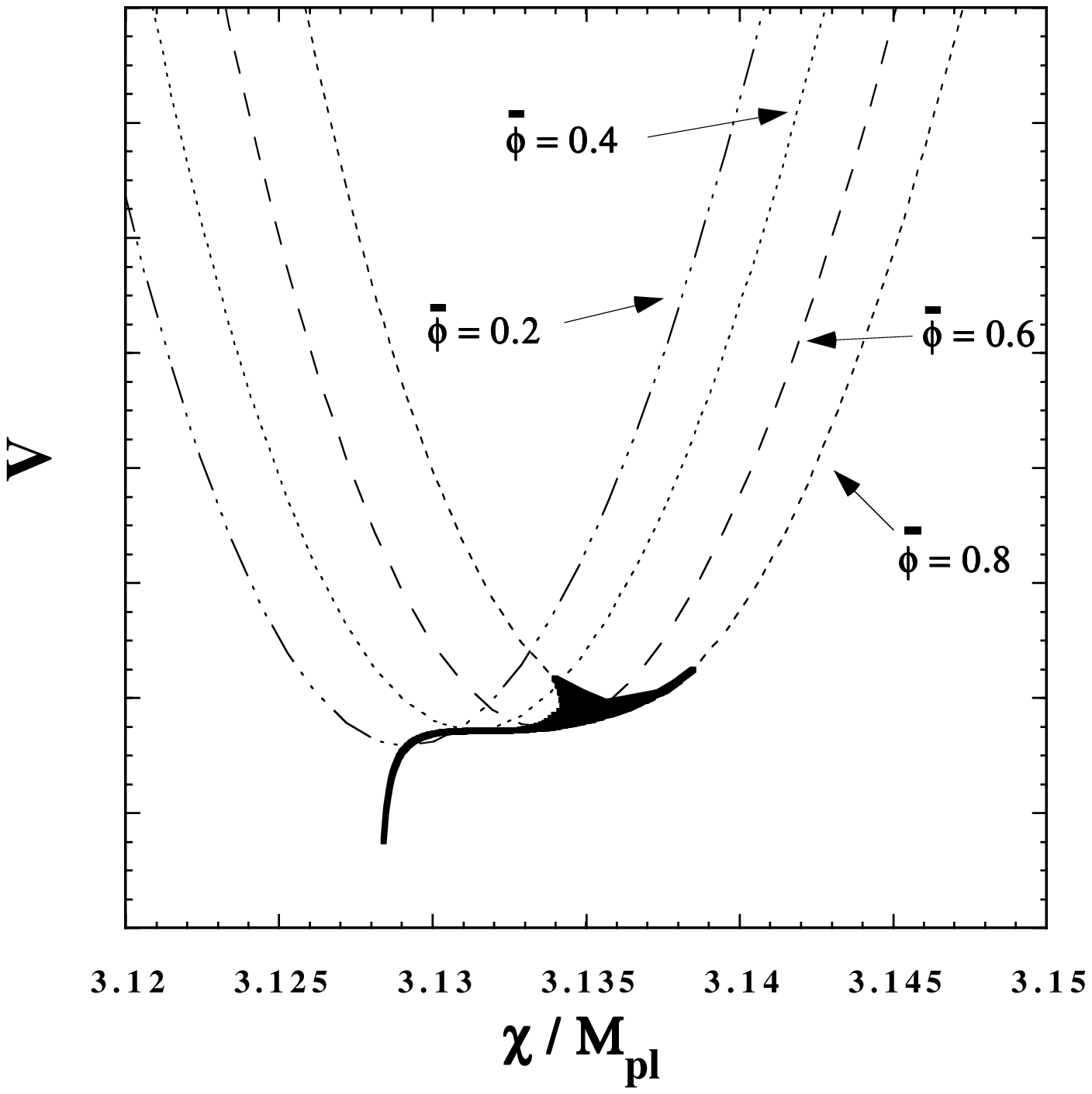}
\par\end{centering}
\caption{\label{fig2} The potential $V(\phi,\chi)$ 
for the model parameters $n=8$, $A_0=1$, 
$\omega_0\equiv b\sigma_0=10.1$,
$\omega_F=9.9951$, $W_0=3.496 \times 10^{-4}$,
$D_0=1.215 \times 10^{-8}$ and $\phi_{\mu}=0.25$.
The solid curves correspond to
the one derived by solving the background equations numerically for
the initial conditions $\phi_i/\phi_\mu=0.8$ and
$\dot{\phi}_i/\phi_\mu=-1.0\times 10^{-10}m$, where 
$m=10^{-7}M_{\rm pl}$ is a mass to normalize time $t$. 
Note that the initial conditions for the field $\chi$ are chosen to satisfy 
the relation (\ref{chire}), i.e., $\chi_i/M_{\rm pl}=3.1385$.
The upper panel shows the potential in terms of the function
of $\phi/\phi_\mu$ for several fixed values of
$\bar{\chi} \equiv \chi/M_{\rm pl}$ (plotted as dotted curves).
The lower panel shows the potential in terms of
the function of $\chi/M_{\rm pl}$ for several fixed values of
$\bar{\phi} \equiv \phi/\phi_{\mu}$. }
\end{figure}

In the upper panel of Fig.~\ref{fig2} we plot the potential
$V(\phi,\chi)$ as a function of $\phi/\phi_\mu$ for the same model
parameters as in Fig.~\ref{fig1}. 
Note that the field $\phi$ is obtained by projecting a two-field trajectory
into the $\phi$ direction.
Since $\eta_{\chi \chi}$ is of order $10^3$ in this case, the mass of the
field $\chi$ is much larger than the Hubble parameter. When
$\chi/M_{\rm pl}>3.141$ the potential does not possess instantaneous
minima in the region $0<\phi/\phi_\mu<1$, but they appear for
$\chi/M_{\rm pl}<3.141$. We require that $\chi/M_{\rm pl}$ is
initially  smaller than 3.160 for $\phi_i/\phi_\mu=0.8$ in order to
avoid that the field evolves toward the forbidden region
$\phi/\phi_\mu>1$.

The numerical simulation in Fig.~\ref{fig2} corresponds to the
initial conditions $\phi_i/\phi_\mu=0.8$ and $\chi_i/M_{\rm
pl}=3.1385$, which satisfy the relation (\ref{chire}). In this case
the field $\phi$ slightly evolves toward larger $\phi$ at the
initial stage, but it soon begins to roll down to instantaneous
minima which move toward smaller $\phi$ with the decrease of $\chi$.
{}From the lower panel of Fig.~\ref{fig2} we find that the field
$\chi$ does not exist at the instantaneous minimum initially (because
of the breakdown of the approximation (\ref{chire})) and then
evolves toward the instantaneous minima with oscillations. The upper
panel of Fig.~\ref{fig2} shows that the instantaneous minima in the
$\phi$ direction disappear for $\chi/M_{\rm pl}<3.130$. This is the
signal of the end of inflation. Thus the field $\chi$ does not
evolve much during inflation in the above case, as required from the
stabilization of the volume modulus. However, as we will see below,
it is of crucial importance to incorporate the dynamics 
of the field $\chi$.

\begin{figure}
\includegraphics[height=3.1in,width=3.1in]{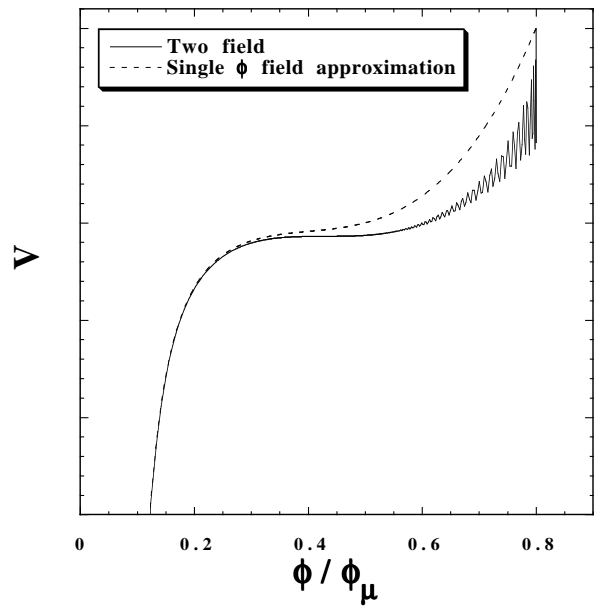}
\caption{\label{fig3} The solid curve represents the potential
obtained by numerically solving the background equations in
two-field system for the same model parameters and initial
conditions as in Fig.~\ref{fig2}.
The dotted curve corresponds to
the potential (\ref{Vstar}) under the single $\phi$
field approximation.}
\end{figure}

In Ref.~\cite{Bau3} the authors reduce the two-field potential
(\ref{potential}) to the single-field one by substituting
the relation (\ref{chire}) for (\ref{potential}):
\ba
\label{Vstar}
V_*(\phi)=V(\phi, \sigma_*(\phi))\,.
\ea
In Fig.~\ref{fig3} we plot the numerically obtained
potential $V$ as a function of
$\phi/\phi_\mu$ to compare with (\ref{Vstar}).
The potential in the two-field system is flatter than
in the case derived under the single $\phi$ field approximation.
This implies that the real two-field system
chooses a trajectory which gives a larger amount of inflation.

\begin{figure}
\includegraphics[height=3.1in,width=3.1in]{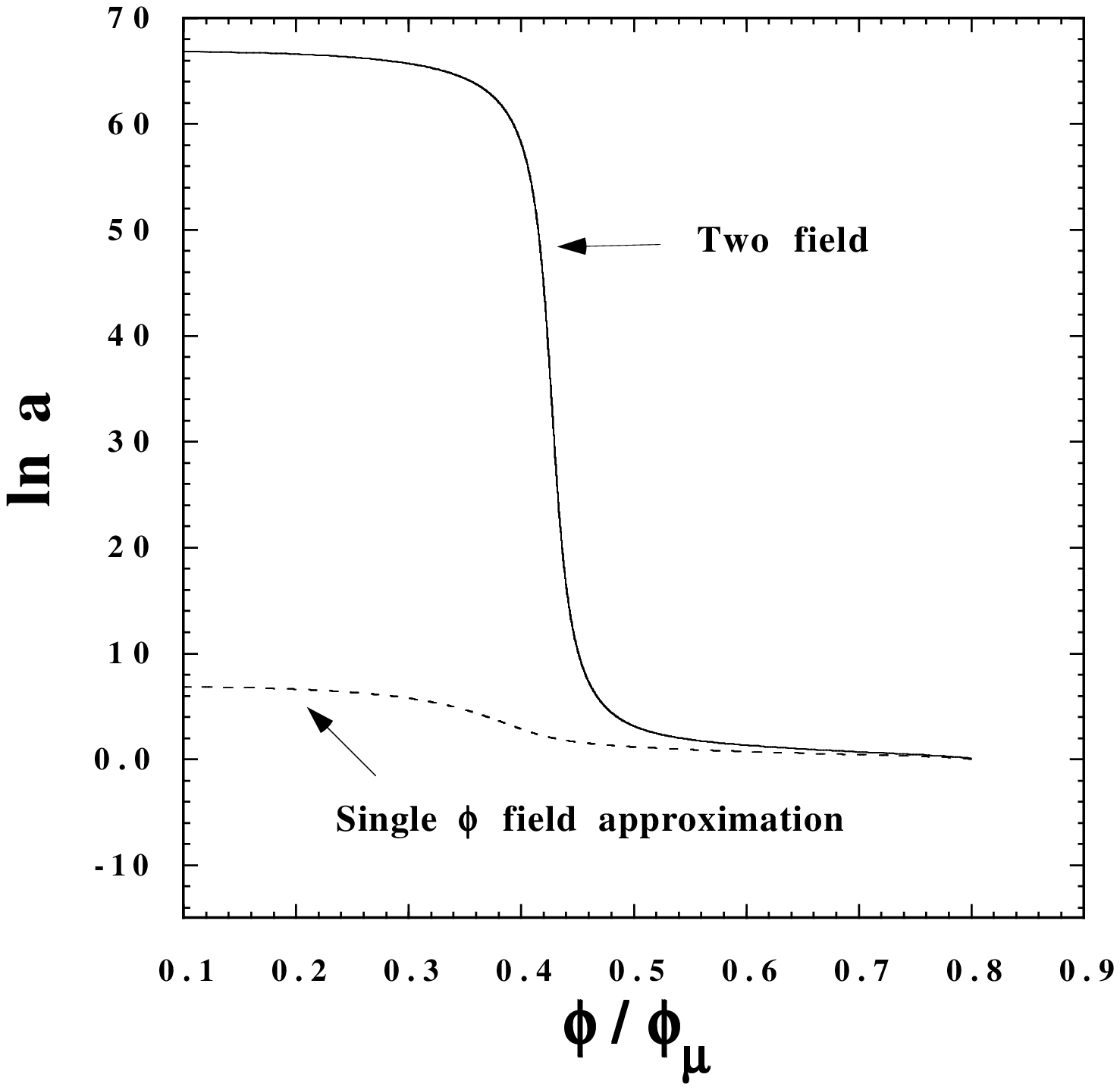}
\caption{\label{fig4} The evolution of the number of e-foldings
in terms of the function of $\phi/\phi_\mu$
for the two-field system (solid curve)
and for the system under the single $\phi$ field approximation
(dotted curve). The model parameters and
initial conditions are the same as in Fig.~\ref{fig2}. }
\end{figure}

In Fig.~\ref{fig4} we show the evolution of the number of e-foldings
${\rm ln }\, a$ in terms of the function of $\phi/\phi_\mu$
with the initial conditions $\phi_i/\phi_\mu=0.8$ and 
$\dot{\phi}_i/\phi_\mu=-1.0\times 10^{-10}m$, where 
$m=10^{-7}M_{\rm pl}$.
In the two-field case inflation occurs around the region
$0.3 \lesssim \phi/\phi_\mu \lesssim 0.5$,
which leads to the number of e-foldings ${\rm ln}\,a=67$ at the end of
inflation. We find that this is not sensitive to the change of initial conditions 
as long as $\phi_i/\phi_\mu$ is larger than 0.5. The change of initial velocities
of scalar fields hardly affects the evolution of the slow-roll regime,  
unless we choose unnaturally large initial velocities.
In the single $\phi$ field approximation, compared to the two-field system,
we obtain a much smaller value of the number of e-foldings: ${\rm ln}\,a=7$.

This discrepancy reflects the fact that the background trajectory
along which the fields evolve is not given by the field $\phi$ but by the
field $\psi$ satisfying the relation
\ba
\label{dpsi}
\dot{\psi}=(\cos \theta) \dot{\phi}+
(\sin \theta) \dot{\chi}\,,
\ea
where
\ba
\label{tanthe}
\tan \theta=\dot{\chi}/\dot{\phi}\,.
\ea
Here $\theta$ characterizes the direction along which the
scalar fields evolve.
If the field trajectory is curved as in our case then we have
$\dot{\theta} \neq 0$.
Since $\dot{s} \equiv -(\sin \theta) \dot{\phi}+
(\cos \theta) \dot{\chi}=0$ from Eq.~(\ref{tanthe}),
the fields do not move to the direction orthogonal to $\psi$.

In order to find out an effective single-field trajectory we need to
obtain the potential $V$ in terms of the function of $\psi$ rather
than $\phi$. If the field $\chi$ is frozen at some particular value
($\chi={\rm const}$), we just need to derive the potential with
respect to $\phi$ because $\psi$ coincides with $\phi$.
The hybrid inflation model with the potential
$V=\frac{\lambda}{4} (\chi^2-M^2/\lambda)^2 +\frac12 g^2 \phi^2
\chi^2+\frac12 m^2 \phi^2$ \cite{hybrid}
falls into this category: inflation occurs
along the line $\chi=0$. However the model (\ref{potential}) gives
a curved background trajectory, which means that the single-field
description in terms of $\psi$ is necessary to understand the dynamics
of inflation correctly.

\begin{figure}
\includegraphics[height=3.1in,width=3.1in]{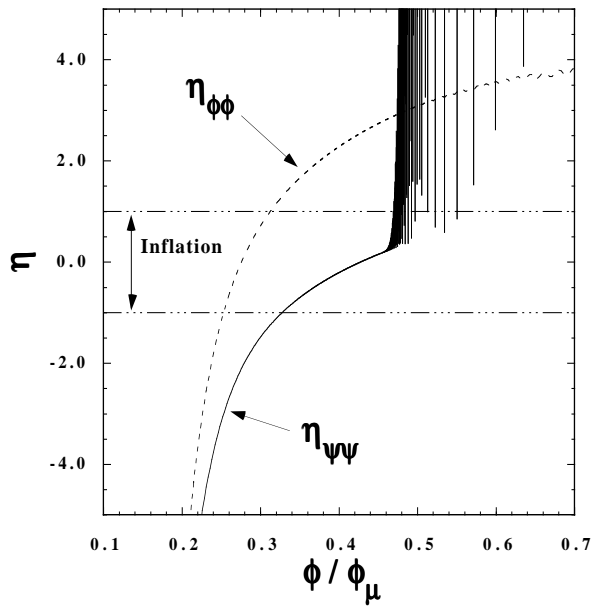}
\caption{\label{fig5} The evolution of the slow-roll
parameters $\eta_{\psi \psi}$ and $\eta_{\phi \phi}$
in terms of the function of $\phi/\phi_\mu$.
The model parameters and initial conditions are chosen as
in the case of Fig.~\ref{fig2}.
The period of inflation is determined by the condition
$|\eta_{\psi \psi}|<1$. }
\end{figure}
\begin{figure}
\includegraphics[height=3.1in,width=3.1in]{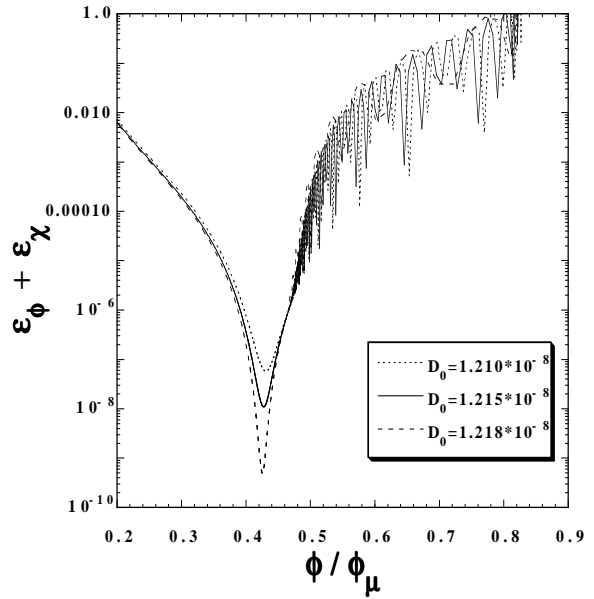}
\caption{\label{fig6} The evolution of $\epsilon_\phi+\epsilon_\chi$
in terms of the function of $\phi/\phi_\mu$ with three different values 
of $D_0$. Other model parameters and initial conditions are chosen as
in the case of Fig.~\ref{fig2}. We find that $\epsilon_\phi+\epsilon_\chi$
are smaller than $10^{-4}$ during inflation ($0.3<\phi/\phi_{\mu}<0.5$).
}
\end{figure}

The mass squared of the field $\psi$ is given by \cite{Gordon}
\ba
\label{Vpsi}
V_{,\psi \psi}=(\cos^2 \theta)V_{,\phi \phi} +
(\sin 2\theta)V_{,\phi \chi}+(\sin^2 \theta)V_{,\chi \chi}\,.
\ea
Then the slow-roll parameter, $\eta_{\psi \psi} \equiv
M_{\rm pl}^2 V_{,\psi \psi}/V$, is
\ba
\eta_{\psi \psi}=(\cos^2 \theta)\eta_{\phi \phi} +
(\sin 2\theta)\eta_{\phi \chi}+(\sin^2 \theta)
\eta_{\chi \chi}\,.
\ea
In Fig.~\ref{fig5} we plot the evolution of the
slow-roll parameters $\eta_{\psi \psi}$
and $\eta_{\phi \phi}$ for the same model parameters
and initial conditions as in Fig.~\ref{fig2}.
The period of the inflationary stage is determined by the
condition $|\eta_{\psi \psi}| <1$ instead of
the condition  $|\eta_{\phi \phi}| <1$.
In fact Fig.~\ref{fig4} shows that inflation occurs for
$0.3 \lesssim \phi/\phi_{\mu} \lesssim 0.5$, which coincides
with the region  given by the condition
$|\eta_{\psi \psi}| < 1$ in Fig.~\ref{fig5}.
The slow-roll parameter $\eta_{\phi \phi}$ is larger than unity
during inflation, which implies that this variable is not suitable
to describe the dynamics of inflation.
In the next section we show that the slow-roll parameter $\eta_{\psi \psi}$,
instead of $\eta_{\phi \phi}$, is crucially important to estimate the spectral
index of scalar metric perturbations.

The scalar fields approach instantaneous minima of the potential so
close that slow-roll parameters $\epsilon_\phi$ and $\epsilon_\chi$
become much smaller than $|\eta_{\psi \psi}|$ during inflation, 
see Fig.~\ref{fig6}.
It is possible to have a
larger total number of e-foldings than the case shown in
Fig.~\ref{fig4} by choosing slightly different values of $D_0$. For
example, when $D_0=1.218 \times 10^{-8}$ instead of $D_0=1.215
\times 10^{-8}$, we find that the total number of e-foldings becomes
${\rm ln }\, a=148$. In this case the field stays at instantaneous
minima of the potential for a longer time.
Then we obtain smaller values of $\epsilon_\phi$ and $\epsilon_\chi$, 
which leads to a larger amount of inflation. 
This behavior is clearly seen in the numerical simulation 
of Fig.~\ref{fig6}.
Inflation ends when the slow-roll parameter
$|\eta_{\psi \psi}|$ becomes larger than unity. Meanwhile if we
choose $D_0=1.210 \times 10^{-8}$ then the total number of
e-foldings is found to be ${\rm ln }\, a=43$, which is not
sufficient to solve horizon and flatness problems. When $D_0=1.220
\times 10^{-8}$ inflation does not end because the fields are stuck
at the local minimum of the potential. These results show how the
number of e-foldings is sensitive to the choice of model parameters.
Thus we require a severe fine-tuning as it is typical 
in the context of D-brane inflation.
In the field equations (\ref{phiequ}) and (\ref{chiequ}) 
we have dropped other kinetic  terms coming from 
the K\"ahler potential which are suppressed by 
${\cal O}(\phi/M_{\rm pl}^2)$. We have carried out 
numerical simulations by taking into account those terms 
and found that the total number of e-foldings is hardly changed (the 
difference of order 0.1). 
Thus the results are insensitive to the inclusion of  
such kinetic terms.

The dynamics of inflation is mainly affected by the changes
of the model parameters $W_0$ and $D_0$. We have run
our numerical code for many other cases by changing $W_0$
and $D_0$ and found that the inflationary dynamics
discussed above is a rather generic feature when the field $\chi$
is massive ($\eta_{\chi \chi} \gg 1$).
The single $\phi$ field approximation using the potential
(\ref{Vstar}) tends to underestimate the total amount
of inflation. We have also considered the parameter regions
in which the $\chi$ mass is smaller than the Hubble parameter
($\eta_{\chi \chi} \lesssim 1$).
In such cases, we numerically find that the field $\chi$ typically 
evolves toward larger values unlike the estimation given in 
Eq.~(\ref{chire}).
In this case the instantaneous 
minima of the potential tend to disappear 
in the region $0<\phi/\phi_{\mu}<1$, which implies
a difficulty to obtain a large number of e-foldings.
Hence in what follows we concentrate on the case where
the mass of the field $\chi$ is much larger than
the Hubble parameter.

\section{Cosmological perturbations}

In this section we discuss the spectra of density
perturbations generated in the model with the
potential (\ref{potential}).
In two-field models of inflation the resulting density
perturbations are generally different from those in
a single-field model because of the presence of isocurvature
(entropy) perturbations \cite{isopapers}
(see Refs.~\cite{Kodama,BTW} for review).
We denote the field perturbations in $\phi$ and $\chi$
as $\delta \phi$ and $\delta \chi$.
Along and orthogonal to the background
trajectory in  two-field space it is convenient to carry
out a field rotation \cite{Gordon}:
\ba
\label{delpsi}
& & \delta \psi \equiv (\cos \theta)\delta \phi +
(\sin  \theta) \delta \chi\,,\\
\label{dels}
& &  \delta s \equiv -(\sin  \theta) \delta \phi+
(\cos \theta) \delta \chi\,,
\ea
where $\theta$ is defined by Eq.~(\ref{tanthe}).
Here $\delta \psi$ and $\delta s$ correspond to adiabatic and
entropy perturbations, respectively.

The perturbed spacetime about the FRW background
is described by the line element
\begin{eqnarray}
\hspace*{-0.25em}\rd s^2 &=& - (1+2A)\rd t^2 +
2a(\partial_iB-S_i)\rd x^i\rd t
\nonumber\\
\hspace*{-0.25em}&& +a^2\left[
(1-2\varphi)\delta_{ij} + 2\partial_{ij}E 
+ h_{ij} \right] \rd x^i \rd x^j,
\end{eqnarray}
where $A$, $B$, $E$, $\varphi$ are scalar metric perturbations and
$h_{ij}$ is the tensor perturbation (we omit to write vector perturbations).
The comoving curvature perturbation ${\cal R}$ and the isocurvature
perturbation ${\cal S}$ are defined by
\begin{eqnarray}
{\cal R}=\varphi+\frac{H \delta \rho}{\dot{\rho}}\,,\quad
{\cal S}=\frac{H(\dot{\phi}\delta \chi-\dot{\chi}\delta \phi)}
{\dot{\phi}^2+\dot{\chi}^2}\,,
\end{eqnarray}
where $\delta \rho$ is the total density perturbation.
Using the field perturbations (\ref{delpsi}) and (\ref{dels})
these are simply expressed as \cite{Gordon}
\begin{eqnarray}
\label{calR}
{\cal R} = \frac{H\delta \psi_\varphi}{\dot{\psi}}\,,\quad
{\cal S}=\frac{H\delta s}{\dot{\psi}}\,,
\end{eqnarray}
where $\delta \psi_\varphi=
\delta \psi+\dot{\psi}\,\varphi/H$.

The each Fourier mode for $\delta \psi_{\varphi}$ and
$\delta s$ satisfies the following equations of
motion \cite{Gordon}
\begin{eqnarray}
\label{adiabaticeom}
\hspace{-1.2em}
& &\ddot{\delta \psi_\varphi}+3H \dot{\delta \psi_\varphi}
+\left[ \frac{k^2}{a^2}+V_{,\psi \psi}-\dot{\theta}^2
-\frac{1}{M_{\rm pl}^2a^3}
\left( \frac{a^3 \dot{\psi}^2}{H}
\right)^{\cdot}\right] \delta \psi_\varphi \nonumber \\
\hspace{-3.0em}& & =2(\dot{\theta} \delta s)^{\cdot}
-2\left(  \frac{V_{,\psi}}{\dot{\psi}}+\frac{\dot{H}}{H}
\right) \dot{\theta}\delta s\,, \\
\label{entropyeom}
\hspace{-1.2em}
& &\ddot{\delta s} + 3H\dot{\delta s} + \left(\frac{k^2}{a^2}
+ V_{,ss} + 3\dot{\theta}^2 \right) \delta s =
\frac{\dot{\theta}}{\dot{\psi}}
\frac{4M_{\rm pl}^2k^2}{a^2}\Psi\,,
\end{eqnarray}
where $k$ is a comoving
wavenumber, $\Psi=\varphi+a^2H (\dot{E}-B/a)$,
$V_{,\psi \psi}$ is defined by (\ref{Vpsi}) and
\begin{eqnarray}
\hspace*{-1.0em}
V_{,ss} = (\sin^2 \theta) V_{,\phi \phi}
-(\sin 2\theta)V_{,\phi \chi}+
(\cos^2 \theta) V_{,\chi \chi}.
\label{Vdd}
\end{eqnarray}
This shows that in the large-scale limit ($k\to 0$)
the entropy perturbation ${\cal S}$ evolves
independently of the curvature perturbation,
while the curvature perturbation ${\cal R}$
is sourced by the entropy perturbation
as long as the trajectory is curved
($\dot{\theta} \neq 0$) in the field space.

The background trajectories we discussed in the previous section
correspond to the cases in which the field $\psi$
is light and the field $s$ is heavy relative to the Hubble parameter
during inflation.
Hence the mass term $V_{,\psi \psi}$
in Eq.~(\ref{adiabaticeom}) is smaller than the order of $H^2$,
whereas $V_{,ss}$ is larger than $H^2$.
When both $V_{,\psi \psi}$ and $V_{,ss}$ are smaller than $H^2$,
it was shown in Ref.~\cite{WBMR,Byr} that
at the Hubble radius crossing ($k=aH$) the cross-correlation
${\cal C}_{{\cal R}S}$ between curvature
and isocurvature perturbations is zero at lowest order
in slow-roll (while it does not vanish at the first order).
In our case ($\eta_{,ss} \gg 1$)
the power spectrum of the entropy perturbation,
${\cal P}_{\delta s} \equiv \frac{k^3}{2\pi^2} |\delta s|^2$, behaves
in the usual way (${\cal P}_{\delta s} \simeq (k/2\pi a)^2$)
for $k^2/a^2 \gg V_{,ss}$, but after the mass term $V_{,ss}$
becomes larger than $k^2/a^2$ the entropy perturbation
decreases more rapidly (${\cal P}_{\delta s} \propto a^{-3}$).
This latter stage occurs even before the Hubble radius 
crossing (i.e., $V_{,ss}>k^2/a^2>H^2$),
which generally leads to the weaker cross-correlation at $k=aH$
relative to the case of two light scalar fields studied
in Refs.~\cite{WBMR,Byr}.

We then neglect the interacting terms on the r.h.s. of Eq.~(\ref{adiabaticeom})
for the modes $k >aH$ and obtain the spectrum
${\cal P}_{\delta \psi_{\varphi *}} \simeq (H_*/2\pi)^2$ at the Hubble
exit at lowest order in slow-roll of the field $\psi$
(in what follows we use the symbol $*$ to represent the quantities
at the Hubble radius crossing, $k=aH$).
Hence from Eq.~(\ref{calR}) the power spectrum of
the curvature perturbation at $k=aH$ is given by
\begin{eqnarray}
{\cal P}_{{\cal R}_*} \simeq
\left( \frac{H^2}{2\pi \dot{\psi}} \right)_*^2
\simeq \left( \frac{V}{24\pi^2 \epsilon
M_{\rm pl}^4} \right)_* \,,
\end{eqnarray}
where
\begin{eqnarray}
\epsilon \equiv \frac{M_{\rm pl}^2}{2}
\left( \frac{V_{,\psi}}{V} \right)^2
\simeq \epsilon_\phi+\epsilon_\chi\,.
\end{eqnarray}
Here the last approximate equality holds under the slow-roll
approximation.  Note that we used the relation
$\dot{\psi}^2 \simeq (2/3)\epsilon V$
and $3H^2 \simeq V/M_{\rm pl}^2$.

One can describe the evolution of perturbations
after the Hubble exit ($k<aH$)
by using a transfer matrix \cite{WBMR}
\begin{equation}
\label{defTransfer}
\left(
\begin{array}{c}
{\R} \\ {\Sc}
\end{array}
\right) = \left(
\begin{array}{cc}
1 & {T}_{\R\Sc} \\ 0 & {T}_{\Sc\Sc}
\end{array}
\right) \left(
\begin{array}{c}
\R_* \\\Sc_*
\end{array}
\right)\,,
\end{equation}
where $T_{{\cal R}{\cal S}}$ characterizes the correlation
between curvature and isocurvature perturbations.
The dimensionless measure of correlation is defined by
\begin{eqnarray}
\label{rcdef}
r_c \equiv \frac{T_{{\cal R}{\cal S}}}
{\sqrt{1+T_{\R\Sc}^2}}\,,
\end{eqnarray}
which is in the range $|r_c| \le 1$ by definition.

If the masses of two scalar fields $\phi$ and $\chi$ are small
relative to the Hubble parameter, it was found in Refs.~\cite{Par} that
the correlation measure $|r_c|$ can be close to the order of 1
in several models of two-field inflation. This comes from the
fact that the entropy perturbation $\delta s$ is not
suppressed after the Hubble radius crossing.
Meanwhile if one of the fields is heavy then the amplitude of $\delta s$
exponentially decreases ($|\delta s| \propto a^{-3/2}$), which
results in a very weak correlation ($|r_c| \ll 1$).
Since we are considering this latter situation for the model
(\ref{potential}), it is a good approximation to neglect the
correlation after the Hubble radius crossing.

Using this property, the power spectrum of the curvature perturbation
at the end of inflation is given by
\begin{eqnarray}
\label{spe2}
{\cal P}_{\cal R} \simeq
{\cal P}_{{\cal R}_*} \simeq
\left( \frac{V}{24\pi^2 \epsilon
M_{\rm pl}^4} \right)_* \,,
\end{eqnarray}
which holds for $|r_c| \ll 1$.
When the correlation is strong, the r.h.s. of Eq.~(\ref{spe2})
is multiplied by the factor $1/(1-r_c^2)$ \cite{WBMR}.
The spectrum index of the curvature perturbation,
$n_{\cal R} \equiv 1+{\rm d} \ln {\cal P}_{\cal R}
/{\rm d} \ln k$, is given by
\begin{eqnarray}
\label{nR}
n_{\cal R} \simeq 1-6\epsilon+2\eta_{\psi \psi }\,,
\end{eqnarray}
where the slow-roll parameters are evaluated at the
Hubble exit.

The tensor perturbation $h_{ij}$ is decoupled from
the scalar perturbation and is frozen after the Hubble
radius crossing. Thus its power spectrum is
given by \cite{WBMR}
\begin{eqnarray}
{\cal P}_{T}={\cal P}_{T_*}
\simeq \frac{2V_*}{3\pi^2 M_{\rm pl}^4}\,,
\end{eqnarray}
with the spectral index
\begin{eqnarray}
n_T \equiv \frac{{\rm d} \ln P_T}{{\rm d} \ln k}=-2\epsilon\,.
\end{eqnarray}
We also obtain the tensor to scalar ratio
\begin{eqnarray}
\label{ratio}
r \equiv \frac{{\cal P}_T}{{\cal P}_{\cal R}}
\simeq 16\epsilon \simeq 16 (\epsilon_\phi+
\epsilon_\chi)\,,
\end{eqnarray}
which is valid for $|r_c| \ll 1$.
In the presence of the correlation the r.h.s. of Eq.~(\ref{ratio})
is multiplied by the factor $(1-r_c^2)$ \cite{Bartolo,WBMR}.

As we mentioned in the previous section, the slow-roll parameters
$\epsilon_\phi$ and $\epsilon_\chi$ for the model (\ref{potential})
are typically very much smaller than 1 because the scalar fields
evolve around instantaneous minima of the 
potential (see Fig.~\ref{fig6}). This implies
that the tensor to scalar ratio given in Eq.~(\ref{ratio}) is much
smaller than one. In the numerical simulation of Fig.~\ref{fig2},
for example, we obtain $r$ of the order $10^{-5}$ on
cosmologically relevant scales. Hence this model satisfies the
present observational upper bound: $r < 0.3$ \cite{Spergel2,Teg}. We
also note that the spectral index $n_T$ of tensor perturbations is
very close to scale-invariant.

Since the model (\ref{potential}) generally
satisfies the relation
$\epsilon \ll |\eta_{\psi \psi}|$ during inflation
except for the vicinity at $\eta_{\psi \psi}=0$, 
the spectral index (\ref{nR}) of
scalar perturbations yields
\begin{eqnarray}
\label{nR2}
n_{\cal R} \simeq 1+2\eta_{\psi \psi}\,.
\end{eqnarray}
It is important to note that $n_{\cal R}$ is determined by
$\eta_{\psi \psi}$ instead of $\eta_{\phi \phi}$.
This reflects the fact that the background trajectory is
not along the $\phi$ direction but along the $\psi$
direction.

The expression (\ref{nR2}) is valid as long as the slow-roll
parameter $\eta_{\psi \psi}$ is smaller than the order of
unity. In the case of Fig.~\ref{fig5}, for example, this can be
used for $\eta_{\psi \psi}<0.2$-0.3 after the fields
almost stop oscillating and enter the inflationary stage.
We have also computed the correlation measure $r_c$ numerically
and have confirmed that $r_c$ is very small
for the modes that crossed the Hubble radius during
slow-roll inflation so that the expression (\ref{nR2}) is valid.

\begin{figure}
\includegraphics[height=3.1in,width=3.3in]{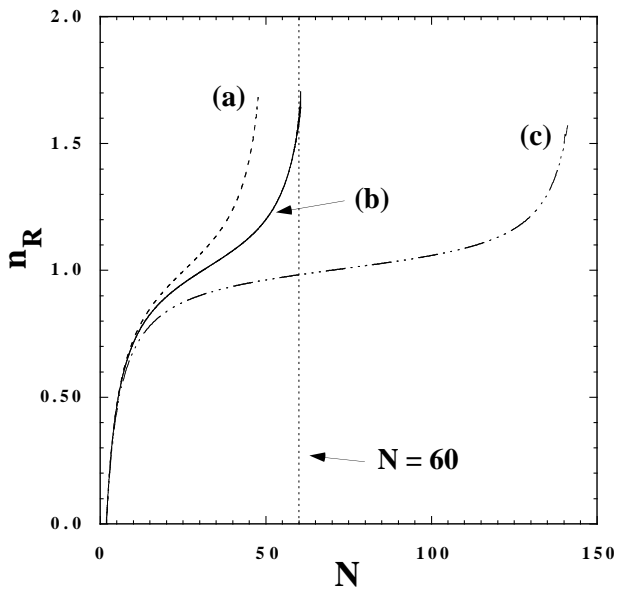}
\caption{\label{fig7} The spectral index $n_{\cal R}$ as
a function of the number of e-foldings $N$ from the end
of inflation.  Each case corresponds to
(a) $D_0=1.213 \times 10^{-4}$,
(b) $D_0=1.215 \times 10^{-4}$, and
(c) $D_0=1.218 \times 10^{-4}$. Other model parameters
are the same as in Fig.~\ref{fig2}.
}
\end{figure}

In Fig.~\ref{fig7} we plot $n_{\cal R}$ in terms of
the function of the number of e-foldings $N$ from the
{\it end} of inflation for several different cases.
For larger $D_0$ the total number of e-foldings increases,
which leads to the change of the curves from (a) to (c).
The curve (b) corresponds to the case in which
the number of e-foldings exceeds the typical COBE
scale value $N=60$, while in the case of the curve (a)
it does not reach cosmologically relevant scales.
In the case (b) we obtain the value $n_{\cal R}
\simeq 1.6$ for $N=60$, which is too large to
satisfy recent observational constraints:
$n_{\cal R}=0.97$-$1.21$ \cite{Kin}.\footnote{Without the
running of the spectral index the constraint on $n_{\cal R}$
is severer: $n_{\cal R}=0.93$-$1.01$.}

It is possible to realize the red-tilted spectrum ($n_{\cal R}<1$)
that satisfies observational constraints if cosmologically relevant
scales are in the negative $\eta_{\psi \psi}$ region.
In the case (c) of Fig.~\ref{fig7}, for example,
we obtain $n_{\cal R} \simeq 0.98$ for $N=60$.
It is clear from Fig.~\ref{fig7} that 
the spectral indices around $N=60$ get smaller 
for larger total number of e-foldings.

\begin{figure}
\includegraphics[height=3.1in,width=3.3in]{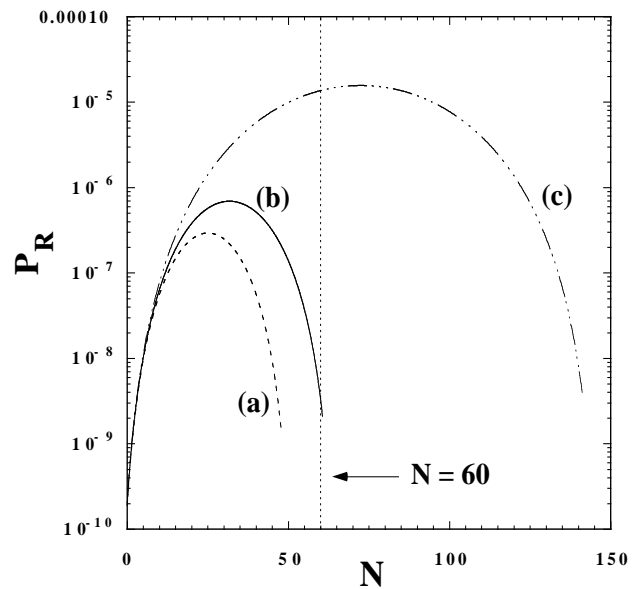}
\caption{\label{fig8} The amplitude of the power spectrum
${\cal P}_{\cal R}$ as a function of the number of e-foldings $N$  
from the end of inflation.  Each case corresponds to
(a) $D_0=1.213 \times 10^{-4}$,
(b) $D_0=1.215 \times 10^{-4}$, and
(c) $D_0=1.218 \times 10^{-4}$. Other model parameters
are the same as in Fig.~\ref{fig2}.}
\end{figure}

The model also needs to satisfy the condition of
the COBE normalization \cite{Spergel2}
\begin{eqnarray}
\label{COBE}
{\cal P}_{\cal R} \simeq 2.4 \times 10^{-9}\,,
\end{eqnarray}
on cosmologically relevant scales observed by COBE. 
In Fig.~\ref{fig8} we find that
in the case (b) this condition can be satisfied around the scale
$N=60$, while the spectral index $n_{\cal R}$ is larger than
observationally allowed values. We obtain the smaller $n_{\cal R}$
consistent with observations in the case (c), but the amplitude
${\cal P}_{\cal R}$ becomes much larger than the value (\ref{COBE}).
The increase of the amplitude reflects the fact
that for larger total number of e-foldings the fields are stuck 
around instantaneous minima
of the potential for a longer time so that $\epsilon$ tends to be
smaller in Eq.~(\ref{spe2}). Thus for the model parameters chosen in
Figs.~\ref{fig7} and \ref{fig8} the model does not satisfy, 
simultaneously, both observational constraints of 
$n_{\cal R}$ and ${\cal P}_{\cal R}$. 

We have tried many other cases and have not found a case in which
both $n_{\cal R}$ and ${\cal P}_{\cal R}$ satisfy observational
constraints. The behavior shown in Figs.~\ref{fig7} and \ref{fig8}
is typical in our model. In order to obtain a nearly scale-invariant
spectrum, we need to adjust that cosmologically relevant scales
($N\sim 60)$ exist in the region around the maximum of the power
spectrum. In this case, however, the amplitude ${\cal P}_{\cal R}$ tends to
be too large to satisfy the COBE normalization. Nevertheless we have
not searched for all parameter spaces; presumably there may be some
regions or isolated points in the parameter space that satisfy both
constraints of $n_{\cal R}$ and ${\cal P}_{\cal R}$ simultaneously. 
However the difficulty to find such viable parameters implies that the model
requires
severe fine-tunings in order to be consistent with observations.

\section{Conclusions}

In this paper we have studied inflation in ``delicate''
D-brane cosmology \cite{Bau2,Bau3} that takes into
account effects of the moduli stabilization.
The presence of the D3-brane in a warped conifold background
gives rise to a correction to the warped four cycle volume,
which leads to a modification to the nonperturbative superpotential
$W$. Hence the total potential $V$ depends upon not only
the inflaton field $\phi$ but also the (canonical) volume modulus field 
$\chi$.

Since the field $\chi$ evolves in two-field space even if the change
is small, we need to study the dynamics of multi-field inflation in an
appropriate way. If the mass of $\chi$ is much larger than the
Hubble parameter, this field quickly evolves toward instantaneous
minima of the potential.
In Ref.~\cite{Bau2,Bau3,KP} the approximate relation (\ref{chire})
with respect to the non-canonical field
$\sigma=\exp(\sqrt{2/3}\,\chi/M_{\rm pl})$ 
was derived under the condition $V_{,\sigma}=0$.
This relation tends to be accurate for smaller
$\phi$, as we have shown in Fig.~\ref{fig1}.
The effective single field potential in terms of $\phi$ can be
obtained by substituting Eq.~(\ref{chire})
for the two-field potential (\ref{potential}).
Numerically we have found that this description
underestimates the total number of e-foldings even by one
order of magnitude, see Fig.~\ref{fig4}.

This discrepancy comes from the fact that the actual background
trajectory is not given by the effective single field $\phi$ but
by the field $\psi$ defined by Eq.~(\ref{dpsi}).
If the scalar fields evolve along a straight line with constant
$\chi$ during inflation, as in the case of hybrid inflation
with the potential
$V=\frac{\lambda}{4} (\chi^2-M^2/\lambda)^2
+\frac12 g^2 \phi^2 \chi^2+\frac12 m^2 \phi^2$, 
the single-field description with respect to $\phi$ is valid
because the $\psi$ direction coincides with the $\phi$
direction. If the trajectory is curved as in the model
(\ref{potential}), one has to study 
the dynamics of inflation along the $\psi$ direction.
In fact in the numerical simulation of Fig.~\ref{fig5}
the period of inflation ($0.3<\phi/\phi_{\mu}<0.5$)
coincides with the one derived by the slow-roll
condition $|\eta_{\psi \psi}|<1$,
but the slow-roll parameter $\eta_{\phi \phi}$
is larger than unity during this epoch.

We also find that the total number of e-foldings is very sensitive
to a slight change of model parameters. This is related to the fact
that the naive single $\phi$ field description grossly
underestimates the total amount of inflation. Although the change of
the field $\chi$ during inflation is typically small, the evolution of
the volume modulus nontrivially affects the dynamics of inflation.
It is interesting to note that in two-field model (\ref{potential})
one can realize a large number of e-foldings to solve horizon and
flatness problems even if it is difficult in single-field context.

We have also evaluated the power spectra of density perturbations
generated in this model.
In two-field models of inflation there exist isocurvature perturbations
in addition to curvature perturbations in general.
When one of the fields is much heavier than the
Hubble rate, the correlation between curvature and isocurvature
perturbations is generally very small.
In this case the power spectrum ${\cal P}_{\cal R}$ of scalar metric perturbation is
estimated by Eq.~(\ref{spe2}) together with the spectral index $n_{\cal R}$
given by Eq.~(\ref{nR2}). It is important to recognize that 
the spectral index is determined by 
the slow-roll parameter $\eta_{\psi \psi}$
instead of $\eta_{\phi \phi}$.
We have found that it is generally difficult to satisfy observational
constraints of both the spectral index and the COBE 
normalization simultaneously.
This comes from the fact that as the spectrum approaches
scale-invariant ($n_{\cal R}=1$) the amplitude ${\cal P}_{\cal R}$
tends to be larger than the COBE normalized value
(${\cal P}_{\cal R} \simeq 2.4 \times 10^{-9}$) on cosmologically
relevant scales, see Figs.~\ref{fig7} and \ref{fig8}.

The model we have studied is in fact ``delicate'' and needs severe
fine-tunings of model parameters to satisfy several constraints
discussed in this paper. The implementation of
corrections due to the D3-brane motion to multi-throat D-brane models
\cite{IT} can possibly improve the situation. It is to be noted that a
viable model of inflation should be followed by a successful
reheating \cite{LKS} which is a challenging problem for D-brane
cosmology at present. The multi-throat models have recently given 
some hopes in this direction, see Refs.~\cite{RH} on the related
theme. These are important issues of string
cosmology which require further investigation.

\section*{ACKNOWLEDGMENTS}
We thank D.~Baumann, A.~Dymarsky, I.~Klebanov, L.~McAllister,
E.~Pajer and  I.~Thongkool for useful discussions
and clarifications. S.\,P. and M.\,S. thank JSPS for financial
supports for their visits to Japan and for kind hospitality in Gunma
National College of Technology, Nagoya University and Tokyo
Institute of Technology.  S.\,P. is supported by DST/JSPS (Grant
No.\,DST/INT/JSPS/Proj-35/2007). 
M.\,S. is supported by DST/JSPS (Grant No.\,DST/INT/JSPS/Proj-35/2007), 
JSPS fellowship (FY2007) and by ICTP
and IUCAA through their associateship programs. 
S.\,T. is supported by JSPS (Grant No.\,30318802).

\end{document}